\documentclass[aps,prl,10pt,preprint,nofootinbib]{revtex4-1}
\usepackage{graphicx}
\usepackage{amsmath}
\usepackage{amssymb}
\usepackage{color}

\begin{document}

\title{Strong field induced electron density tide in high harmonic generation from solids}

\author{Liang Li$^{1,\text{a}}$, Yinfu Zhang$^{1,}$\footnote{These authors contribute equally to this work}}
\author{Pengfei Lan$^{1}$}\email{Corresponding author: pengfeilan@hust.edu.cn}
\author{ Tengfei Huang$^{1}$, Xiaosong Zhu$^{1}$, Chunyang Zhai$^{1}$, Ke Yang$^{1}$, Lixin He$^{1}$, Qingbin Zhang$^{1}$, Wei Cao$^{1}$}
\author{Peixiang Lu$^{1,2,3}$}\email{Corresponding author: lupeixiang@hust.edu.cn}

\affiliation{
 $^1$Wuhan National Laboratory for Optoelectronics and School of Physics, Huazhong University of Science and Technology, Wuhan 430074, China\\
 $^2$Hubei Key Laboratory of Optical Information and Pattern Recognition, Wuhan Institute of Technology, Wuhan 430205, China\\
 $^3$CAS Center for Excellence in Ultra-intense Laser Science, Shanghai 201800, China
}

\date{\today}

\begin{abstract}
In strong laser fields, the electron density in solids can show up tidal motions caused by the laser ``pulling'' on the electrons, which forms an electron density tide (EDT). However, the strong field processes in solids are always explained by the single-active-electron (SAE) model and the fluctuation of background electrons is neglected previously. Here, we demonstrate the strong field induced EDT effect and propose a model for revealing its role in high harmoinc generation (HHG) from solids. We show that the EDT effect induces an additional polarization current beyond the SAE approximation. It gives a new mechanism for HHG and leads to new anisotropic structures, which are experimentally observed with MgO. Our experiment indicates that the EDT effect becomes more obvious with increasing the laser intensity. Our work establishes the bridge between the HHG and the microscopic EDT in solids induced by the strong laser field, which paves the way to probe the ultrafast electron dynamics.
\end{abstract}


\maketitle

The ocean tide and atmospheric tide are typical phenomena due to the periodic gravitation variation. In strong laser fields, the electron density can show up similar tidal motions caused by the laser ``pulling'' on the electrons. Such electron motion forms a electron density tide (EDT) and should have non-negligible influence in the interaction of intense laser pulses and solids. In this work, we concentrate its role in the high harmonic generation (HHG) from solids. 

Recently, HHG in solids has attracted much attention for its potential applications as a source of the extreme utraviolet radiation \cite{Kim2017,Garg2018,McDonald2017,Vampa2017}. It also provides opportunities for crystallographic analysis and probing the electronic properties \cite{Zaks2012,Hohenleutner2015,Langer2016,Silva2018,Luu2018,Yu2018,Luu2015,Vampa2015,Lanin2017}, opening a new field of attosecond physics in solids. For all these applications, it is very important to clearly understand the physics underlying solid HHG. Current investigations are mostly based on the single-active-electron(SAE) approximation models \cite{Vampa2014,Higuchi2014,Luu2016,Osika2017,Mengxi2017,Mengxi2015,Ikemachi2017,Mengxi2016,Dejean2017,Golde2008,Bian2017,Osika2017,Li2018}. By modeling the electron motion in the energy bands, the main features of solid HHG, e.g., the cutoff law and the multi-plateau structure of the spectrum, can be explained. The multielectron effects, e.g., the EDT and scattering between electrons, are neglected or simply described by phenomenologically introducing the dephasing time. Although the ab initio simulations can include the multielectron effect \cite{dftsim1,dftsim2}, the underlying mechanisms are buried in the wave functions, making it difficult to understand the mechanisms. Indeed, recent reports on the orientation dependence of harmonic yields and its relation to the crystal structure \cite{You2016,Shima2017} indicate that the underlying physics of solid HHG is not fully understood.

In this Letter, we demonstrate that the so-far overlooked EDT induced by strong laser field plays an important role in solid HHG. We propose a theoreatical model to reaveal the EDT effect and our model shows that the EDT effect can be attributed to the multielectron scattering. It leads to an oscillating dipole near the cores and then induced an additional polarization current beyond the SAE models. This results in new anisotropic structures of the HHG yield. To demonstrate this effect, we experimentally investigate HHG in MgO with laser fields varying from linear to circular polarization. The experimental results show that, with increasing the laser intensity, the HHG yield exhibits different anisotropic structures as a function of laser ellipticity and crystal orientation. These experimental observations are out of reach of the previous SAE models \cite{Mengxi2017}. In contrast, they are remarkable signatures of the EDT effect.

\begin{figure}[!t]
	\includegraphics[width=16cm]{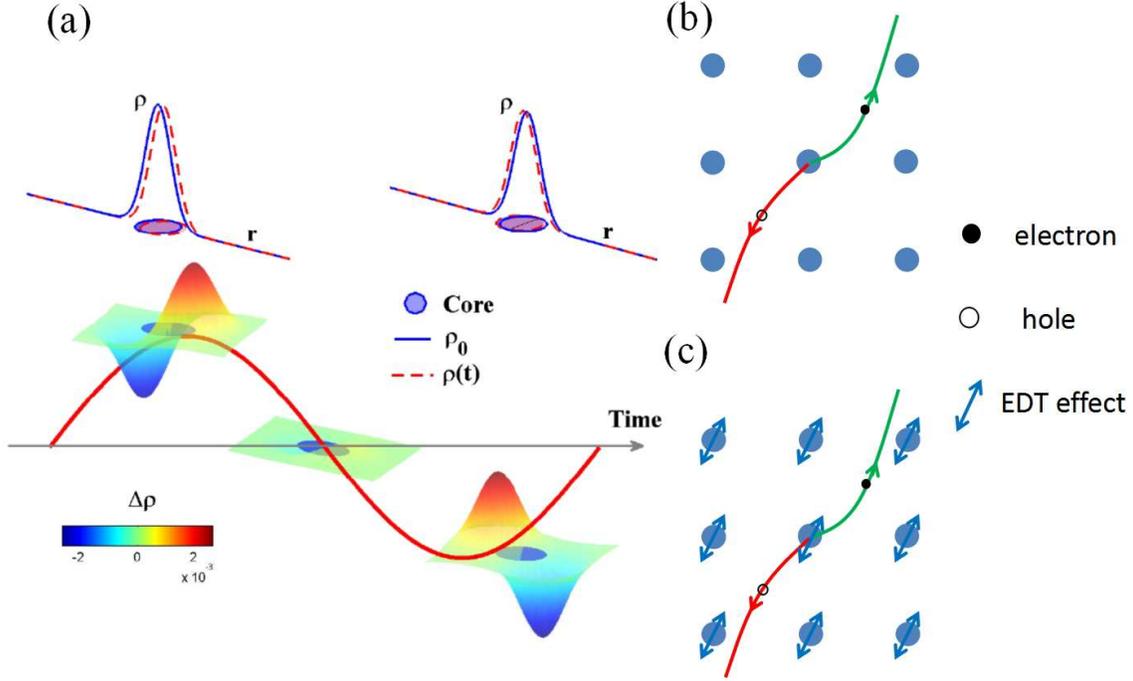}
	\caption{(a) Upper plots: Sketch of the strong field induced EDT along the polarization direction of the laser pulse. In the strong laser field, the electron density $\rho$ is modified compared to the initial density $\rho_0$. Bottom plots: the variation of the charge density $\Delta\rho=\rho(t)-\rho_{0}$ under a linearly polarized laser pulse. The simulation is performed with TDDFT for a model lattice (see supplementary materials \cite{SM}).  $\Delta\rho$ is negligible around the zero point of the electric field. While in the first half cycle, more electrons are pulled to the right side by the strong laser field, e.g., $\Delta\rho>0$ at the right side. In the next half cycle, more electrons are pulled to the left side when the electric field changes its sign. (b) The sketch of the trajectory for SAE models. During the acceleration, the electron (hole) is moving under the frozen effective potential. (c) The sketch of the trajectory with EDT effect. In the strong laser field, the electron density changes back and forwards, resulting in the EDT. The electron(hole) is moving under the time-dependent potential of EDT-cores.}\label{fig1}
\end{figure}

To explain the EDT, we start from the ab initio Hamiltonian and apply the generalized Coulomb gauge,
\begin{align}\label{Hamiltonian}
H = \sum_{j}\frac{[\textbf{p}_{j}+\textbf{A}(\textbf{r}_{j},t)]^2}{2}+\frac{1}{2}\sum_{j}V_{j,J}(\textbf{r}_{j},\textbf{R}_{J})+\frac{1}{2}\sum_{j,k\neq j}V_{j,k}(\textbf{r}_{j},\textbf{r}_{k})+\frac{1}{2}\sum_{J,K\neq J}V_{J,K}(\textbf{R}_{J},\textbf{R}_{K}).
\end{align}
Here the Born-Oppenheimer approximation is applied to separate the electronic motion from the nuclear motion. The nuclear movement is much smaller than that of the electron and is neglected. We use capital notations (J, K, R) for the nuclei and the lower-case notations (j, k, r) for electrons. By adopting the dipole approximation to the laser field and mean-field approximation for the crystal systems, the electron dynamics is governed by the Kohn-Sham (KS) equations \cite{Runge1984}
\begin{align}\label{TDDFT}
i\frac{\partial}{\partial t}\Psi_{i}(\textbf{r},t)=[\frac{\textbf{p}^{2}}{2}+v_{\text{eff}}(\textbf{r},t)+H_{I}(t)]\Psi_{i}(\textbf{r},t)
\end{align}
with effective potential
\begin{align}\label{v_efft}
v_{\text{eff}}(\textbf{r},t) = \int\frac{\rho(\textbf{r}',t)}{|\textbf{r}-\textbf{r}'|}d\textbf{r}'+v_{\text{n}}(\textbf{r})+v_{\text{xc}}[\rho(\textbf{r},t)]
\end{align}
The effective potential consists of the electron-electron, electron-nuclear, and exchange-correlation terms. $H_{I} = \frac{1}{2}(\textbf{A}\cdot\textbf{p}+\textbf{p}\cdot\textbf{A})$ is the laser-electron interaction. The nuclear-nuclear interaction is neglected here and will not influence the discussion. Previous models employ a time-independent $v_{\text{eff}}$ with stationary electron density distribution $\rho_{0}$ that satisfies the time-independent KS equations. Then the system is modeled by considering the single electron motion, i.e. SAE, driven by the laser field \cite{Vampa2014,Higuchi2014,Luu2016,Osika2017,Mengxi2017,Mengxi2015,Ikemachi2017,Mengxi2016,Dejean2017,Golde2008,Bian2017,Osika2017,Li2018}. The HHG process is explained as: (as shown in Fig. \ref{fig1}(b)) the electron is ionized from the parent core and then accelerated by the laser field, accompanied by the generation of induced currents and high harmonics under the frozen potential $v_{\text{eff}}[\rho_{0}]$.

However, high harmonics are generated with a strong field that is comparable to the electron-bounding potential and the bound electrons are generally much less localized in solids compared with those in gases. Therefore, the bound electrons in solids can be pulled back and forth under the strong laser field. In Fig. \ref{fig1}(a), we show the change of the valence charge density $\rho(\textbf{r},t)-\rho_{0}$ simulated by the time-dependent density functional theory \cite{Andrade2015} (detalis are shown in supplementary materials \cite{SM}). One can see obvious fluctuation of the electron density near the core (a totality of nuclei and inner electrons called the EDT-core in the following) even when the ionization is low ($<0.1\%$). In this case, the effective potential also changes following the EDT. Then, the electron dynamics will be influenced when it is scattered by the EDT-core and differnt polarzation current can be induced compared with the SAE models, which can paly a remarkable role in solid HHG. 

\begin{figure}[!t]
	\includegraphics[width=16cm]{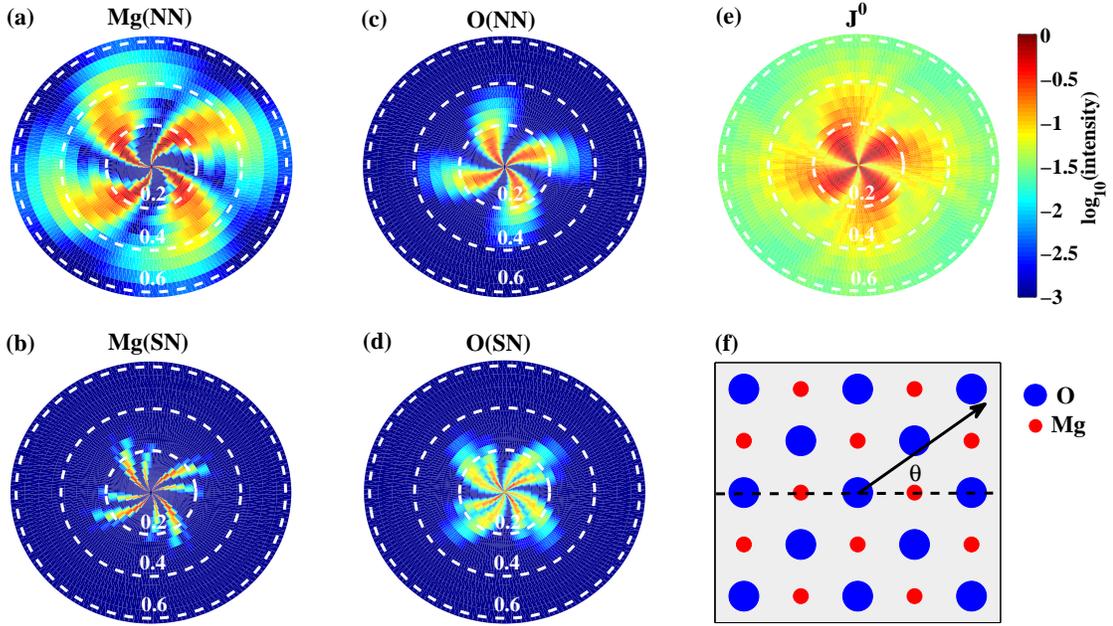}
	\caption{Orientation dependence of the 17th harmonic yield under different ellipticities contributed by the electron density fluctuation effect of (a) the nearest-neighbor Mg, (b) the second-neighbor Mg, (c) the nearest-neighbor O, (d) the second-neighbor O, and (e) the normal current $\textbf{J}^{0}$. (f) The sketch of a $\langle001\rangle$ cut MgO crystal. $\theta$ indicates the orientation angle between the crystal and the main axis of the laser polarization. The radius in (a)-(e) corresponds to the ellipticity, which ranges from $\varepsilon=0$ to $\varepsilon=0.63$. The white dashed lines in Figs. \ref{fig3}(a), (b), and (c) corresponds to $\varepsilon=0.2$, $\varepsilon=0.4$, and $\varepsilon=0.6$. The harmonic yield are normalized to 1 in these figures.}\label{fig2}
\end{figure}

To model the EDT effect, we involve $H_{s} = v_{\text{eff}}(\textbf{r},t)-v_{\text{eff}}(\textbf{r},t)\lvert_{t=0}$ as a perturbation term of the Hamiltonian, and then the wavefunction can be express by (see the supplementary materials \cite{SM}),
\begin{align}\label{wavefunction}
|\Psi_{i}(t)\rangle = |\Psi_{i}^{0}(t)\rangle+U_{0}(t,t_{0})\int^{t}_{t_{0}} d\tau U^{\dagger}_{0}(\tau,t_{0})[-iH_{s}(\tau)]U_{0}(\tau,t_{0})|\Psi_{i}^{0}(t_{0})\rangle.
\end{align}
$|\Psi_{i}^{0}(t)\rangle$ is the wavefunction without considering the EDT effect. The second term $|\Psi_{i}^{s}(t)\rangle=U_{0}(t,t_{0})\int^{t}_{t_{0}} d\tau U^{\dagger}_{0}(\tau,t_{0})[-iH_{s}(\tau)]U_{0}(\tau,t_{0})|\Psi_{i}^{0}(t_{0})\rangle$ is induced by the EDT, which will lead to additional polarization currents. The total current can be expressed as $\textbf{J}(t)  = \sum_{i}\langle\Psi_{i}(t)|\textbf{p}|\Psi_{i}(t)\rangle= \sum_{i}\left[\langle\Psi_{i}^{0}(t)|\textbf{p}|\Psi_{i}^{0}(t)\rangle+\langle\Psi_{i}^{0}(t)|\textbf{p}|\Psi_{i}^{s}(t)\rangle+\langle\Psi_{i}^{s}(t)|\textbf{p}|\Psi_{i}^{0}(t)\rangle\right]$. This can be divided into two parts, $\textbf{J}^{0}(t)$ and $\textbf{J}^{s}(t)$, which we evaluate in the framework of quantum trajectories (see the supplementary materials \cite{SM}),
\begin{align}\label{Current}
\textbf{J}^{0}(t) &= \sum_{i}\langle\Psi_{i}^{0}(t)|\textbf{p}|\Psi_{i}^{0}(t)\rangle \nonumber \\ &=\sum_{t',\textbf{k}_{0}\in BZ}T_{cv}(t')e^{iS_{cv}(t,t')}\textbf{p}_{cv}(\textbf{k}(t))e^{-(t-t')/T_{2}}+c.c.   \nonumber \\
\textbf{J}^{s}(t) &= \sum_{i}\left[\langle\Psi_{i}^{0}(t)|\textbf{p}|\Psi_{i}^{s}(t)\rangle+\langle\Psi_{i}^{s}(t)|\textbf{p}|\Psi_{i}^{0}(t)\rangle\right] \nonumber \\ &=\sum_{t',\textbf{k}_{0}\in BZ}\left[\sum_{L}e^{-i\textbf{k}(t)\cdot\textbf{R}_{L}}\xi(\textbf{R}_{L},t)\right]T_{cv}(t')e^{iS_{cv}(t,t')}\textbf{p}_{cv}(\textbf{k}(t))e^{-(t-t')/T_{2}}+c.c.
\end{align}
Here, only two bands are considered, i.e., one valence band (VB) and one conduction band (CB). The intraband current is omitted due to its minor contribution to the high harmonics above the band gap. $T_{cv}(t')$ is the ionization rate, and $\textbf{p}_{cv}(\textbf{k})$ is the transition matrix element between the CB (denoted as ``$c$'') and the VB (denoted as ``$v$''). $S_{cv}(t,t_{i}) = \int_{t_{i}}^{t}\epsilon_{g}(\textbf{k}(\tau))d\tau$ is the dynamical phase. $\epsilon_{g}$ is the band gap, and $\textbf{k}(t) = \textbf{k}_{0}+\textbf{A}(t)$ with $\textbf{k}_{0}$ belonging to the first Brillouin zone (BZ). We also include the dephasing factor by introducing an attenuation term $e^{-(t-t_{i})/T_{2}}$ to describe the isotropic part of the scattering effects \cite{McDonald2017,Vampa2017}. The first term $\textbf{J}^{0}$ is the normal current obtained with previous SEA models \cite{Li2018,Mengxi2017} excluding the EDT effect. The orientation dependence of the corresponding HHG is determined by the frozen mean field $v[\rho_{0}]$. For MgO, this term leads to a four-arm structure of the orientation dependence of HHG as shown in Fig. \ref{fig2}(e) and also Ref. \cite{Mengxi2017}. The second term $\textbf{J}^{s}$ is contributed by the strong field induced EDT. This term contains a factor $\xi(\textbf{R}_{L},t)$ corresponding to the electron-electron scattering, which can be approximately evaluated by the overlap between the ionized electron and the bound electron wave packet near the core, i.e., $\xi(\textbf{R}_{L},t)\sim e^{-(\textbf{r}(t)-\textbf{R}_{{L}})^2/a_{\text{core}}^2}$, where the subscript ``core'' corresponds to Mg or O and $\textbf{R}_{L}$ is location of the cores. $a_{\text{Mg}}$ and $a_{\text{O}}$ are the sizes of Mg and O cores, respectively. It means that, due to the EDT, the ionized electron will be scattered by the bound electrons when it passes through the EDT-cores. The scattering modifies the current and leads to $\textbf{J}^{s}$, which depends on $\textbf{R}_{L}$. This effect will give rise to new structure of HHG in contrast to the normal currents $\textbf{J}^{0}$. In addition, the ellipticity dependence of HHG will also be influenced by the EDT effect.

To demonstrate the role of the EDT, we discuss HHG from MgO in the laser fields varying from linear to circular polarization. Figure \ref{fig2} shows the yield of the 17th harmonic as a function of crystal orientation and laser ellipticity as predicted by our model. The model also enables us to separate the current and HHG into the contributions from the scattering with different EDT-cores. As shown in Fig. \ref{fig2}, all the spectrograms show a spiral structure. For the signal of the nearest-neighbor (NN) Mg and O, see Figs. \ref{fig2}(a) and (c), there are four paddles, which corresponds to the four-fold symmetry of the atomic configuration. In the linear polarization case, the signal of Mg(NN) peaks along the bonding directions, e.g. 0$^\circ$, 90$^\circ$, while the signal of O(NN) peaks in the middle of each quadrant, e.g. 45$^\circ$, 135$^\circ$. With increasing the ellipticity, the peaks of harmonic yields are shifted clockwise away from the bonding directions. Specifically, the Mg(NN) peak moves from 0$^\circ$ to about 45$^\circ$, and the O(NN) peak moves from 45$^\circ$ to about 25$^\circ$. Wider paddles can be seen in larger ellipticity until saturation at $\varepsilon\approx0.5$. In addition, as shown in Figs. \ref{fig2}(b) and (d), the signal of the second-neighbor (SN) Mg and O exhibits 8 paddles. In the linear polarization case, the harmonic yield peaks at the orientations along the direction from the origin to Mg(SN) and O(SN). In contrast, as shown in Fig. \ref{fig2}(e), the HHG from the normal current $\textbf{J}^{0}$ exhibits only 4 paddles peaked at 45$^\circ$ and 135$^\circ$. We conclude that there are two remarkable signatures representing the EDT effect: a new anisotropic structure (8 paddles vs 4 paddles) and different ellipticity and orientation dependence of HHG (Figs. \ref{fig2}(a)-(d) vs Fig. \ref{fig2}(e)), which can be attributed to the $\textbf{R}_{L}$-dependent terms that is only involved in $\textbf{J}^{s}$.

\begin{figure}[!t]
	\includegraphics[width=16cm]{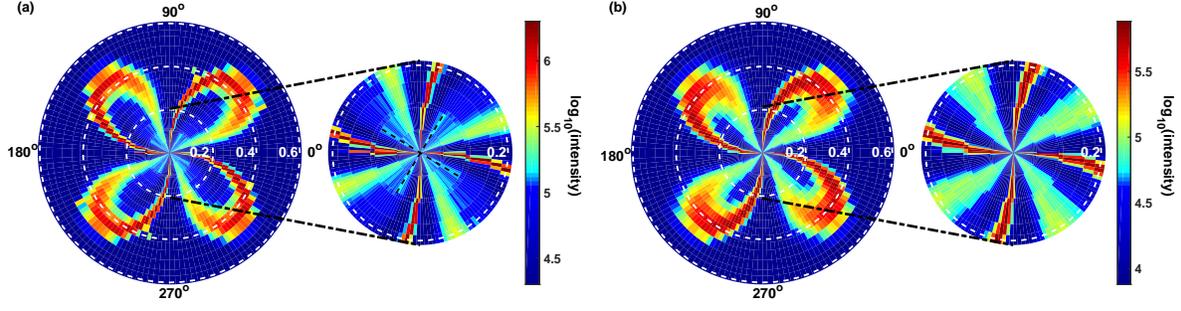}
	\caption{The yield of the 17th harmonic as a function of the crystal orientation and laser ellipticity observed in experiments. The laser energy is (a) 70 $\mu$J, and (b) 40 $\mu$J. The laser wavelength is 1300 nm and pulse duration is about 100 fs. The insets are the results in the low ellipticity region.}\label{fig3}
\end{figure}

\begin{figure}[!t]
	\includegraphics[width=16cm]{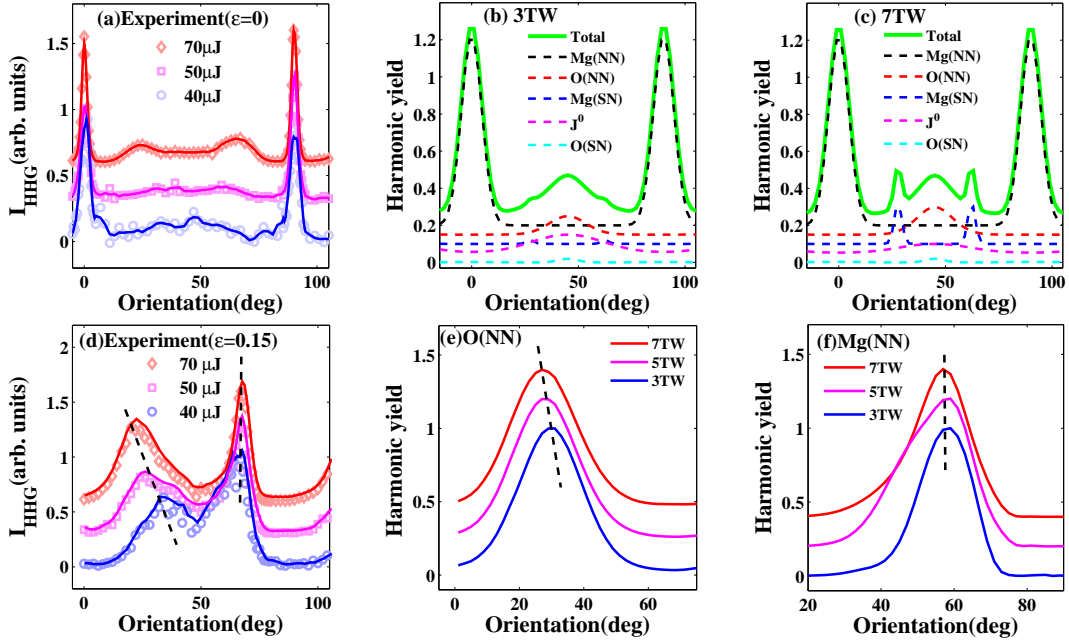}
	\caption{Top row: the orientation dependence of the 17th harmonic yield in the linearly polarized pulse. (a) The experimental results with different laser power. The simulated results with laser intensity (b) 3 TW/cm$^{2}$, and (c) 7 TW/cm$^{2}$. Bottom row: the orientation dependence of the 17th harmonic yield ($\varepsilon=0.15$). The lines are vertically shifted for clarity. (d) The experimental results with different laser power. The lines are vertically shifted for clarity. (e) and (f) Simulation results for HHG yields contributed by the O(NN), and Mg(NN). The results are normalized to 1 and vertically shifted for clarity.}\label{fig4}
\end{figure}

To verify the strong field induced EDT, we perform the experiment by using a near infrared laser. High harmonics are generated by focusing a near-infrared laser on the 300-$\mu$m-thick, 001-cut MgO crystal at normal incidence. The near-infrared laser is produced from an optical parametric amplifier (TOPAS-Prime-Plus) pumped by an 800-nm, 35-fs, 1kHz laser (Coherent, Astrella-USP-1K). The maximum pump energy is 5 mJ and the maximum output energy is about 600 $\mu$J for the signal and 500 $\mu$J for the idle pulses. The wavelength of the near-infrared laser can be adjusted from 1.2 $\mu$m to 2.6 $\mu$m. The pulse duration measured with an autocorrelator is about 100 fs for the signal pulse. The near-infrared laser is focused to a spot of 90-$\mu$m using a lens. We control the laser ellipticity by changing the angle between a half-wave plate and a quarter-wave plate system. The major axis of the elliptical polarization is fixed during our experiments. High harmonics are detected by a homemade flat-field soft x-ray spectrometer consisting of a flat-field grating ($300$ grooves mm$^{-1}$) and a slit. The high harmonics passing through the slit are dispersed by the grating and imaged onto the microchannel plate (MCP) fitted with a phosphor screen. The image on the screen is detected by a CCD camera. 

As shown in Fig. \ref{fig3}(a), the HHG yield shows a four-leaf clover structure and a weak cross structure near the origin (see the inset for $\varepsilon\le0.1$). One can see that the ellipticity and orientation dependence of the harmonic yield shows clear dependence on the crystal structure and laser intensity. For lower driving laser power (40 $\mu$J), a similar four-leaf clover structure is observed. However, the cross structure near the origin gradually disappears with decreasing the power. Note that the cross structure is not seen in previous works \cite{You2016} either.

Comparing with the numerical results in Fig. \ref{fig2}, one can see that the experimental observations show clear discrepancies from the simulated results of normal currents $\textbf{J}^{0}$, and this phenomenon becomes more obvious with increasing the laser intensity. By comparing the experiments and simulations, one can identify the contributions from different EDT-cores. Due to the four-fold symmetry of the spectrogram, we only discuss the results in the first quadrant. We can find that the upper half part (from $45^\circ$ to $90^\circ$ in the first quadrant) of the four-leaf clover structure is mainly contributed by Mg(NN), while the bottom part (from $0^\circ$ to $45^\circ$) is mainly contributed by both O(NN) and $\textbf{J}^{0}$. The cross structure is shown only by considering the contributions of Mg(SN).

For clarity, in Figs. \ref{fig4} (a)-(c), we show the lineout of the orientation-dependent harmonic yields in linearly polarized pulses with different intensities. One can see that the normal current $\textbf{J}^{0}$ only contributes a wide peak near 45$^\circ$. The peaks at 0$^\circ$ and 90$^\circ$ can be attributed to the contribution of Mg(NN). With higher laser intensity, two additional peaks at 30$^\circ$ and 60$^\circ$ appear (a similar phenomenon is also observed with different laser wavelengths, see the supplementary mateials \cite{SM}), which corresponds to the cross structure of Mg(SN) shown in Fig. \ref{fig3}(a). These two peaks disappear at low intensity. These results indicate that the contributions of the EDT induced currents become more obvious with increasing laser intensity, and the contribution of Mg(NN/SN) is dominant to that of O(NN/SN).

Next, we discuss the orientation-dependent harmonic yield in elliptically polarized laser fields [see Figs. \ref{fig4} (d)-(f)]. By comparing the numerical and experimental results, one can attribute the wide peak at small angle $\sim30^\circ$ to the contribution of O(NN) and the narrow peak at 60$^\circ$ to Mg(NN). As shown in Figs. \ref{fig4}(e), the wide peak shifts from 30$^\circ$ to 25$^\circ$ with increasing the laser power. In contrast, the shift of the narrow peak is negligible, which is consistent with the experimental results in Fig. \ref{fig4}(d). Moreover, the contributions of SN EDT-cores are less obvious than those of NN EDT-cores in the elliptical polarization case. This can be phenomenologically understood by the electron trajectories: when we increase the ellipticity of the laser field, the electrons follow trajectories with larger extension radius. Then, the electron is rescattered by the NN EDT-cores before it reaches the SN EDT-cores, and this leads the decrease of contributions of the SN EDT-cores.  

Our work demonstrates the strong field induced EDT effect, which has been so far neglected in previous investigations of solid HHG. The new anisotropic structures of orientation and ellipticity dependence of HHG yield from MgO clearly indicate the remarkable role of the EDT. Comparing the theoretical and experimental results, the relation between the atomic configuration and the spectral structure of HHG is revealed. Our work establishes a bridge between the microscopic  dynamics and HHG signal, and suggests a potential approach to measure the valence electron density and the field induced charge migration inside the crystals. Moreover, it also suggests a new level of controlling the electronic properties of solids with strong laser fields, for example, modifying the nonlinear polarization and the microdynamical responses of the solids. 

This paper was supported by the National Key R\&D
program (2017YFE0116600). National Natural
Science Foundation of China (NSFC) (No. 11934006, 91950202, 11627809, 11874165); Numerical simulations presented
in this paper were carried out using the High Performance
Computing experimental testbed in SCTS/CGCL. We acknowledge valuable discussions with Prof. Manfred Lein.

\end{document}